\documentclass[prd,twocolumn,showpacs,preprintnumbers,amsmath,amssymb,nofootinbib]{revtex4}
\usepackage{graphicx}\usepackage{dcolumn}% Align table columns on decimal point
\usepackage{bm}% bold math
\usepackage{amssymb}\usepackage{amsmath}\usepackage{amsthm}\usepackage{latexsym}\newcommand{\be}{\begin{equation}}\newcommand{\ee}{\end{equation}}\newcommand{\bea}{\begin{eqnarray}}\newcommand{\nn}{\nonumber}\newcommand{\eea}{\end{eqnarray}}
\begin{document}
\title{Field equations from a surface term}
\author{Thomas P.~Sotiriou\footnote[2]{sotiriou@sissa.it} and Stefano Liberati\footnote[3]{liberati@sissa.it}}
\affiliation{SISSA/ISAS, via Beirut 2-4, 34014, Trieste, Italy and INFN, Sezione di Trieste}

\begin{abstract}
 As is well known, in order for the Einstein--Hilbert action to have a well defined variation, and therefore to be used for deriving field equation through the stationary action principle, it has to be amended by the addition of a suitable boundary term. It has recently been claimed that, if one constructs an action by adding this term to the matter action, the Einstein field equations can be derived by requiring this action to be invariant under active transformations which are normal to a null boundary. In this paper we re-examine this approach both for the case of pure gravity and in the presence of matter. We show that in the first case this procedure holds for more general actions than the Einstein-Hilbert one and trace the basis of this remarkable attribute. However, it is also pointed out the when matter is rigorously considered the approach breaks down. The reasons for that are thoroughly discussed.
\end{abstract}


\maketitle

\section{Introduction}

 Einstein's first derivation of the homonymous field equations was obtained by looking for generally covariant equations that would generalize the Poisson equation ({\em i.e.}~the equation relating the gravitational field with its source, the matter, in Newtonian dynamics). At the same time Hilbert presented a derivation of these equations from an action, using the principle of stationary action (we prefer this term instead of the more common one, ``least action principle'', since stationary points may generally not be true minima). This has caused some historical debate about who should be given credit for the introduction of the dynamical equations of General Relativity (GR) \cite{meh}. In any case, Hilbert's derivation has become standard due to its elegance and its connection with classical field theory.

In a variational approach to gravity, what specifies the resulting theory is the choice of the gravitational action.
The most common one is to use the so called Einstein--Hilbert action, an integral over the invariant four-volume of the Ricci scalar (scalar curvature). As is well known, the Ricci scalar depends not only on the first derivatives of the metric, but also on the second derivatives. Therefore, one might not expect the variation of the Einstein--Hilbert action with respect to the metric to lead to second order field equations. However, the structure of the action is such that it can be split into two parts: one which depends on the first derivatives of the metric and one that includes all the terms that depend on the second derivatives \cite{lan}. This second part has the form of a four-divergence and so it leads to a surface term. 

The presence of the surface term is not a special characteristic of the Einstein--Hilbert action. One can easily construct a more general action from scalar quantities depending on the metric whose variation with respect to the  latter will lead to surface terms (for example by adding the Gauss--Bonnet invariant, higher order terms in the scalar curvature, etc.). In general, surface terms are neglected in derivations of the field equations from an action. However, the formal way to avoid them is actually to subtract them from the initial action \cite{wald,haw,york}. Recently the importance of surface terms was discussed in \cite{paddy,paddy3}, where it was claimed that the Einstein field equations (plus an undetermined constant) can be derived with an action constructed by the surface term part of the Einstein--Hilbert action and the matter action. The main idea of this derivation is to consider accelerated observers with a horizon. For these observers the horizon will be the physical boundary on which the surface term should be evaluated. Requiring that the action mentioned before should be invariant under an active transformation in the direction of the normal to the horizon (translations of the horizon normal to itself) leads to the field equations. It was claimed in \cite{paddy3} that this result seems to be an indication of an intrinsic holographic nature of gravity, as all the information about  the dynamics of the four dimensional spacetime  appears to be encoded on the three dimensional boundary.

  The requirement for diffeomorphism invariance is often identified in the literature with ``background independence'' of the theory or with the principle of general covariance and observer independence. In fact there has been a historical debate about this issue \cite{nort} and the discussion about the physical consequences of any of the above statements is still open \cite{gul}. 
An established fact is that diffeomorphism invariance is a necessary requirement for a background independent theory, even though it may not be a sufficient one (see \cite{gul} for an enlightening discussion), and includes the notion of observer independence, {\emph i.e.} the fact that the laws of physics are the same for {\it any} observer. Therefore, if we want to construct a background independent gravitational theory, like General Relativity, then the field equations of this theory should be diffeomorphism invariant. At the level of the action, this requirement is expressed through the invariance of the action leading to these equations.

 In a previous version of this paper we attempted to generalize the approach  presented in \cite{paddy,paddy3,paddy2} in order to include more general actions and boundaries \cite{v1}. However in the present paper, we re-examine this procedure and the attempted generalization and we show that even though the results of \cite{paddy,paddy3,paddy2} are valid both for General Relativity and for more general actions when pure gravity is concerned, the same procedure to derive cannot be followed in order to derive field equations in the presence of matter. 

We start by reviewing the standard derivation of the Einstein field equation from the Einstein--Hilbert action. 
We proceed by considering in section \ref{dif} an action that depends only on the metric and will lead to second order field equations under standard metric variation and studying its behaviour under active transformations (diffeomorphisms). It is shown that in the case of pure gravity field equations can be derived by requiring the invariance of the bulk gravitational action under a special class of diffeomoprhisms. The basis for this fact is fully analyzed. In section \ref{surf} we present equivalent results starting from an action built out of the surface term of the gravitational action. We then show in section \ref{matter} that due to the invariance of the matter action field equations including matter cannot be consistently derived through the approach described in \cite{paddy,paddy3}. Finally, we comment on whether gravity can be considered intrinsically holographic and on the remarkable feature that one can derive field equations for pure gravity from both bulk and surface terms using diffeomorphism invariance. Our conclusions are stated in section \ref{concl}.

\section{Field equations}
\label{FE}

The Einstein--Hilbert action is
\be
S_{\rm EH}=\frac{1}{2 \kappa}\int_U d^4 x \sqrt{-g}\, R.
\ee
where $U$ denotes a region of the spacetime manifold $M$, $g$ is the determinant of the metric $g_{\mu\nu}$ defined on $M$, $R$ is the Ricci scalar and $\kappa=8 \pi G$.
Varying this action with respect to the metric we get
\be
\label{veh} \delta S_{\rm EH}=\frac{1}{2 \kappa}\int_U d^4 x\sqrt{-g}\,G_{\mu\nu}\delta g^{\mu\nu}+\delta S^{\rm EH}_{\textrm{S}},
\ee
where $\delta S^{\rm EH}_{\textrm{S}}$ is a surface term. Using the fact that the metric is kept fixed on the boundary during the variation, $S^{\rm EH}_{\textrm{S}}$ can take  the following form
\begin{equation}
S^{\rm EH}_{\textrm{S}}=-\frac{1}{\kappa}\int_{\partial U} d^3 x \sqrt{|h|}\, K,
\end{equation}
where $\partial U$ denotes the boundary of $U$, $h$ is the determinant of the induced metric on $\partial U$ and $K$ is, as usual, the trace of the extrinsic curvature of that boundary \cite{wald}. We can define the stress-energy tensor (SET) in the usual way, $
T_{\mu\nu}\equiv-2(-g)^{-1/2}\delta {\cal L}_{\rm M}/\delta g^{\mu\nu}$,
where ${\cal L}_M$ is the matter Lagrangian density. $S_{\rm M}$ will be used to denote the matter action. If we try to form the full action as 
$S=S_{\rm EH}+S_{\rm M}$,
this action will not lead to the Einstein field equation, as expected, due to the presence of the surface term in $S_{\rm EH}$. What is more, $S$ will not even be functionally differentiable at the solutions of the field equations (see p.~451 of \cite{wald}). The surface term cannot be set to zero, since this would require not only fixing the metric on the boundary, as usual, but also fixing its derivatives, an assumption which is too restrictive \cite{wald,haw,york}. The solution is to simply subtract this term from the initial gravitational action and form a new one
\be
\label{b}
{S}^{\prime}_{\rm EH}=\frac{1}{2 \kappa}\left[\int_U d^4 x \sqrt{-g}\, R+2\int_{\partial U} d^3 x \sqrt{|h|}\, K\right],
\ee
 the variation of which will not lead to this unwanted surface term \footnote{ Remarkably, even though Hilbert did not notice the need for a surface term, Einstein did: already in 1916 \cite{einstein1}, he used 
action (\ref{b}) in the derivation of the field equations. Noticeably, this was almost six decades before action (\ref{b}) was  
rigorously established.
Moreover, revealing his deep insight, he also pointed out, earlier in  
the same year, that the gravitational action does not have to be
necessarily built out of a generally covariant scalar \cite{einstein2}. This is exactly what one needs to realize in order to use action (\ref{b}).}.  Note that this approach actually applies to more general actions, see for example \cite{bunch} for an extension to the Gauss--Bonnet case. So we shall now consider formally the case of a general action and
for simplicity we shall deal with Lagrangians and Lagrangian densities instead of actions whenever possible.

 Let $L_{\rm grav}(g)$ be gravitational Lagrangian constructed only from a symmetric metric $g_{\mu\nu}$; ${\cal L}_{\rm grav}=\sqrt{-g}L_{\rm grav}(g)$ will be the corresponding Lagrangian density. The variation of the latter with respect to the metric gives
\be
\label{lg}
\delta {\cal L}_{\rm grav}=\sqrt{-g}B_{\mu\nu}\delta g^{\mu\nu}+\delta {\cal L}_{\textrm{S}},
\ee
where $\delta {\cal L}_{\rm S}$ will be the part of the variation of the Lagrangian that can be considered as a surface term. Let us identify the rest of the Lagrangian density as the bulk term, {\em i.e.}
\be
\label{lB}
\delta {\cal L}_{\rm B}:=\delta {\cal L}_{\rm grav}-\delta {\cal L}_{\rm S}=\sqrt{-g}B_{\mu\nu}\delta g^{\mu\nu},
\ee
In analogy with the standard treatment presented above, we shall identify this bulk term as the correct one to be used for the derivation of the field equation. So, once
 we consider also the matter fields, the total  bulk Lagrangian density will be
\be
\label{tot1}
{\cal L}_{\textrm{tb}}:={\cal L}_{\rm B}+{\cal L}_{\rm M}.
\ee
Varying with respect to the metric leads to
\be
0=\delta{\cal L}_{\rm B}+\delta{\cal L}_{\rm M}=\sqrt{-g}\,(B_{\mu\nu}-8\pi G\, T_{\mu\nu})\delta g^{\mu\nu},
\ee
and therefore to the field equations
\be
\label{f}
B_{\mu\nu}=8\pi G\, T_{\mu\nu}.
\ee
When $L$ is chosen to be the Einstein--Hilbert Lagrangian, $B_{\mu\nu}=G_{\mu\nu}$  and eq.~(\ref{f}) reduces to the standard Einstein equations, as expected. 

 The previous derivation seems quite straightforward but one has to note that there is a non-trivial step that we bypassed: the fact that the variation of ${\cal L}$ can be split into a bulk term and a surface term (total divergence), as in eq.~(\ref{lB}), does not necessarily imply that one can separate ${\cal L}$ into two parts ${\cal L}_{\rm B}$ and ${\cal L}_{\textrm{s}}$.  In fact, this additionally requires that the non-vanishing part of the surface term can be written as a total variation. If this is not possible one cannot avoid the unwanted surface term in the variation of ${\cal L}$ by redefining the gravitational Lagrangian density that is to be used in eq.~(\ref{tot1}). Therefore in these cases, deriving the field equations by extremizing an action (principle of stationary action) does not seem to be a feasible task. 

Let us elaborate on this point.
Let us consider the case of second order gravity, {\it i.e.}~Lagrangians that lead to second order field equations, like those just mention above. In this case the only variable to be fixed on the boundary is the metric. Thus, if there is no term that one can add into ${\cal L}$ so that the variation of the resulting action will not have non-vanishing surface terms, then the extremization of the action cannot be carried out and the derivation of the field equations via the principle of stationary action cannot be rigorously accomplished. It is, therefore, questionable if any Lagrangian without this attribute would be useful in forming a gravitational theory based on the principle of stationary action.  Hence the requirement of dealing with an action separable into a bulk and a surface part does not arise from our specific treatment, but from the fact that we need to have an action that can be used to derive field equations through standard metric variation. There are indeed several Lagrangians that are known to belong to this class ({\it e.g.}~the Einstein--Hilbert \cite{york} and Gauss--Bonnet types \cite{bunch}). 

The validity of the above statement seems unambiguous for Lagrangians leading to second order field equations. Let us now consider the more complicated case of higher-order Lagrangians. When it comes to higher-order gravity, as correctly stated in \cite{barth}, more degrees of freedom should be fixed on the boundary than those related to the constancy of the metric there. The extra quantities that one chooses to fix are related to the auxiliary variables of the canonical formulation of the theory, so this choice is physically meaningful, even though in a purely classical approach  it might not be obvious. One can, therefore, understand that the choice of the auxiliary variable to be fixed will drastically affect the form of the surface terms that need to be added, if any. Due to this fact, the form, or even the presence of a surface term in higher-order gravity is ambiguous. Thus we will only concentrate here on the second order case.

\section{Field equations and diffeomorphism invariance: bulk action}
\label{dif}

 We have seen in the previous section that the field equations for the case of ``second order'' gravity can be derived by a suitably defined bulk action 
\be
\label{ba}
S_{\rm B}=\frac{1}{2\kappa}\int_U d^4 x \sqrt{-g}\, L_{\rm B},
\ee
where $L_B$ is the Lagrangian that corresponds to the Lagrangian density ${\cal L}_{\rm B}$.
Let us now examine the behaviour of this action under the active transformation $x^{\mu}\rightarrow x^\mu+\zeta^\mu$ \footnote{Such a transformation should be more precisely characterized as a diffeomorphism \cite{wald} but it is commonly referred to in the literature as an  active transformation, so we shall  use here interchangeably these two terminologies. Note that often diffeomorphism are also referred to as infinitesimal or active coordinate transformations (\emph {e.g.~}\cite{paddy}). However, we will avoid using these terms here since during a diffeomorphism the coordinate system remains the same.}. The metric will transform in the following way: $g^{\mu\nu}\rightarrow g^{\mu\nu}+\delta_\zeta g^{\mu\nu}$, where $\delta_\zeta g^{\mu\nu}=\nabla^\mu \zeta^\nu+\nabla^\nu \zeta^\mu$ (which is equal to minus the Lie derivative of the metric $\mathsterling_\zeta g^{\mu\nu}$ along $\zeta^\mu$ as defined in \cite{wald}). 
If we impose the condition
\be
\label{cond11}
\delta_\zeta g^{\mu\nu}|_{\partial U}=0,
\ee
 then  $S_{\rm B}$ will transform in the following way
\be
\label{fe1}
\delta_\zeta S_{\rm B}=\frac{1}{\kappa}\int_U d^4 x \,\sqrt{-g}B_{\mu\nu}\nabla^\mu \zeta^\nu,
\ee
and integrating by parts one gets
\bea
\label{var1}
\delta_\zeta S_{\rm B}&=&-\frac{1}{\kappa}\int_U d^4 x \sqrt{-g}\left[\nabla^\mu B_{\mu\nu}\right]\zeta^\nu+ \label{fe2}\\
& &+\frac{1}{\kappa}\int_U d^4 x\sqrt{-g}\nabla^\mu (B_{\mu\nu} \zeta^\nu).\nn
\eea
Notice that condition (\ref{cond11}) is needed in order to drop surface terms that include $\delta_\zeta g^{\mu\nu}$ and cannot be avoided otherwise (see for example \cite{wald} for the case of the Einstein-Hilbert action)\footnote{ This issue was also mentioned to us by S.~Gao and H.~Zhang in private communication.}. 

Demanding the invariance of the bulk action under diffeomorphisms and using Stokes theorem in the second line of eq. (\ref{var1}) we obtain
\bea
\label{12}
&&\int_U d^4 x \left[\sqrt{-g}\,(\nabla^\mu B_{\mu\nu}) \zeta^\nu\right]-\\&&-\int_{\partial U} d^3 x\left[ \sqrt{|h|}\,(B_{\mu\nu})\zeta^\nu n^\mu\right]=0,\nn
\eea

This equality is composed of two integrals; one over the volume and one over the surface, which have different dependence on the vectors $\zeta^\nu$ and $n^\mu$. 
Hence the two term have to vanish separately.

Imposing the bulk term in the above expression to be zero is nothing more than requiring that $B_{\mu\nu}$ should satisfy a generalized contracted Bianchi identity 
\be
\label{gbianchi}
\nabla^\mu B_{\mu\nu}=0.
\ee

The vanishing of the second term is however non trivial.  In the standard textbook discussion about diffeomorphism invariance it is assumed that $\zeta^\mu|_{\partial U}=0$. This assumption is adequate to justify the validity of eq.~(\ref{cond11}) and at the same time it implies that the integral over the boundary in eq.~(\ref{12}) vanishes. So all one derives from standard diffeomorphism invariance in the contracted Bianchi identity. But, do we have an alternative choice for the generator of our diffeomorphism, $\zeta^\nu$?

As mentioned, the transformations that we are considering are diffeomorphisms,  {\it i.e.}~a mapping of $U$ into itself.  During this procedure we cannot cross the boundary surface of course and the fields should remain unaffected there. This is the reason why in most textbooks $\zeta^\mu$ is assumed to vanish on the boundary ($\zeta^\mu|_{\partial U}=0$). However, there is a class of diffeomorphisms that is fully consistent with the demand that the boundary is not to be crossed without $\zeta^{\mu}$ having to vanish there: those parallel to the boundary, {\it i.e.~}transformations by a vector which is normal to the normal to the boundary, $n^{\mu}$. 
 For a vector that does not vanish on the boundary condition (\ref{cond11}) takes the form
\be
\label{cond1}
\delta_\zeta g^{\mu\nu}|_{\partial U}=[\nabla^\mu \zeta^\nu+\nabla^\nu \zeta^\mu]|_{\partial U}=0,
\ee
which implies that the generator of the diffeomorpism should also satisfy the Killing equation on the boundary in order to leave the field unaffected there. Therefore, if $\zeta^\mu$ is tangential to the boundary and satisfies eq.~(\ref{cond1}) it fulfills the criteria for a diffeomorphism under which the gravitational action should be invariant \footnote{We are discussing the case of pure gravity in which the metric is the only field.}.

 However, one easily notices that condition (\ref{cond1}) is not only a condition on the generator $\zeta^\mu$ but also a condition on the metric, or more precisely on the form of the metric on the boundary. Enforcing this condition for an arbitrary boundary\footnote{The principle of stationary action or diffeomorphism invariance are always applied and should hold for any compact region $U$ with a boundary $\partial U$. To clarify this point let us recall a rigorous formulation of the principle of stationary action: For every measurable set $U\subset M$ in spacetime $M$ one can write the action functional as $S\left[U,\phi\right]$, where $\phi$ collectively denotes the fields that are to be varied. It is to be understood that there is one action functional for each $U$. The stationary action principle is then formulated as $\delta_\phi S\left[U,\phi\right]=0$ for any $U$ and consequently for any $\partial U$. } would be equivalent to requiring that the metric admits a Killing vector and therefore a symmetry. If the metric has such a global symmetry, then condition (\ref{cond1}) is satisfied for an arbitrary boundary. (An example would be the Schwarzchild metric and the boundary can be any three dimensional hypersurface tangential to the Killing vector that acts as a generator.) But assuming {\emph a priori} that the metric has a symmetry is not really an option if one wants to examine the behaviour of the gravitational action without making a mention to the solution of the resulting field equations.

Notice though, that condition (\ref{cond1}) is a condition on the boundary  and therefore only requires a local symmetry. Such a condition is obviously satisfied on a Killing horizon. One can imagine for example a Schwarzchild black hole present in a region of a general spacetime with no global symmetries. On the horizon of this black hole this condition is trivially satisfied. Therefore if one considers a region $U$ with boundary $\partial U$ which is a Killing horizon then eq.~(\ref{12}) should hold and apart from the contracted Bianchi identity one gets
\bea
\label{13}
\int_{\partial U} d^3 x\left[ \sqrt{|h|}\,B_{\mu\nu}\xi^\nu \xi^\mu\right]=0.
\eea
where $\xi^\mu$ now denotes the special $\zeta$ which is a null vector normal (but also tangential) to the null boundary, which now is implicitly a Killing horizon. 

In general for a curved metric one cannot globally define a Killing horizon passing through each point of spacetime. So it might seem that one cannot deduce that the above equations hold everywhere. However, on can still use the equivalence principle and, along the line of~\cite{Jacobson}, consider a small region around each point $p$ of spacetime to be flat. Hence one can always define a local Rindler horizon through this point.  The part of spacetime beyond this Rindler horizon can then  be considered as instantaneously stationary and there will be an approximate Killing vector that generates boosts orthogonal to this horizon.  One can then ask $\xi^\nu$ to be equal to this Killing vector on the local Killing horizon and zero everywhere else including the rest of the boundary. Recall that the local Killing horizon will not be globally defined but one can still use the technique just described and turn the integral over the boundary in eq.~(\ref{13}) to an integral over the local Killing horizon. This appears to be the line implicitly followed in  \cite{paddy,paddy3,paddy2} as well. 

With the above prescription we can now ask the above equation to hold for any local Rindler observer in spacetime. Therefore, one can consider  $\xi^\mu$ arbitrary, apart from being equal to the generator of a local Rindler horizon on the boundary $\partial U$. Eq.~(\ref{13}) then leads to the equation
\be
\label{f2}
B_{\mu\nu}-\Lambda g_{\mu\nu}=0
\ee
where the undetermined constant comes from the fact any scalar depending on metric times the metric can be added in the integrand of eq.~(\ref{13}) without contributing to it, but this scalar turns out to be a constant if the generalized contracted Bianchi identity is invoked.  In this sense one can claim that diffeomorphism invariance, if viewed within the broader perspective presented here,  can lead, apart from the contracted Bianchi identity, also to the gravitational field equations when pure gravity is concerned. 

\section{Field equations and diffeomorphism invariance: surface action}
\label{surf}

Let us now discuss the case in which a derivation of the field equations is attempted starting from a surface action $S_{\textrm{S}}$ which, in analogy with the standard Einstein--Hilbert action case, we will define to be the quantity which we have to subtract from $S$ in order to recover $S_{\rm B}$ 
\be
\label{ssdef}
S_{\textrm{S}}:=-\frac{1}{2\kappa}\int_{U}d^4 x{\cal L}_{\textrm{S}},
\ee
and $S_{\rm grav}:=S_{\rm B}-S_{\textrm{S}}$, (note that the minus in (\ref{ssdef}) makes this definition consistent with eq.~(\ref{lB})).  

For any Lagrangian $L$ which is a generally covariant scalar, one can write~\cite{paddy}
\begin{eqnarray}
\label{5}
\delta_\zeta S=\int_U d^4x \delta_{\zeta}{\cal L}=-\int_{U} d^4 x \sqrt{-g}\nabla_\mu (\zeta^\mu L).
\end{eqnarray}
which, when applied to ${\cal L}_{\rm grav}$, leads to
\be
\label{scal}
\delta_\zeta S_{\rm grav}=-\frac{1}{2\kappa}\int_{\partial U} d^3 x\sqrt{|h|}\, L_{\rm grav}\zeta^\mu n_\mu.
\ee
As we discussed previously, $\zeta^{\mu}$ will either have to vanish on the boundary or it will have to be parallel to it, $\zeta^\mu n_\mu=0$, in order to map the boundary into itself.  In the latter case we shall still have to enforce condition (\ref{cond1}),  but even without any mention to it we can still show that 
$0=\delta_\zeta S_{\rm grav}=\delta_\zeta S_{\rm B}-\delta_\zeta S_{\textrm{S}}$. Therefore, we can deduce that
\be
\label{gc2}
\delta_\zeta S_{\textrm{B}}=\delta_\zeta S_{\rm S},
\ee

 Thus the requirement the that the bulk gravitational action is invariant under the specific class of diffeomorphisms described in the previous section leads to the same requirement for the surface action $S_{\rm S}$.  Let us now try to impose this requirement.  First of all we need to compute the variation of $S_{\textrm{S}}$.  In order to do this let us recall that by definition $\delta {\cal L}_{\textrm{S}}=\delta {\cal L}_{\rm grav}-\delta {\cal L}_{\rm B}$, so that using eqs.~(\ref{fe1}) and (\ref{5}) we get
\bea
\delta_\zeta S_{\textrm{S}} &=& \frac{1}{2\kappa}\int_U d^4 x\sqrt{-g}\,\left[\nabla^\mu(L_{\rm grav} \zeta_\mu)+ 2B_{\mu\nu}\nabla^\mu \zeta^\nu\right]\nn\\
&=&\frac{1}{\kappa}\int_{\partial U} d^ 3 x \sqrt{|h|}\, \left[\left(B_{\mu\nu}+\frac{1}{2}L_{\rm grav} g_{\mu\nu}\right) \zeta^\nu n^\mu \right]-\nn\\
& &-\frac{1}{\kappa}\int_U d^4 x \sqrt{-g}\,\nabla^\mu (B_{\mu\nu}) \zeta^\nu=0
\eea
which again should hold for any $U$ and $\partial U$ provided that the latter is a Killing horizon and hence $\zeta^\nu n^\mu|_{\partial U}=\xi^\nu\xi^\mu$.
The requirement that separately the bulk and the surface terms vanish allows us to recover the results of the previous section, i.e~the generalized Bianchi identity (\ref{gbianchi}) and the  field equation
\be
\label{f3}
B_{\mu\nu}-\Lambda g_{\mu\nu}=0.
\ee

So, in summary, by just considering the case of pure gravity and we have learn that
\begin{itemize}
\item One cannot in general derive the field equations from the requirement of diffeomorphism invariance of the bulk or surface action for gravity. If one does the standard assumption that $\delta_\zeta g_{\mu\nu}=0$ is enforced via the requirement $\zeta |_{\partial U}=0$, then no surface term remains and all one gets is the generalized Bianchi identity.
\item  However, if  one does not assume that $\zeta^\mu |_{\partial U}=0$ but enforces instead (\ref{cond1}) plus the requirement that $\zeta$ is parallel to the boundary, the result is a different special class of diffeomorphisms that the gravitational action has to be invariant under, since they do map a compact region $U$ in itself. Even though considering such a class seems to be a restriction on the form of the metric (e.g. the presence of global Killing horizons) this can be circumvented by considering local Rindler horizons as part of the boundary and suitably restricting the generator $\zeta^\mu$. Using the invariance of either the bulk or the surface gravitational action one can indeed derive field equations for pure gravity which  hold throughout spacetime, since every point in $U$ can be considered as a point of some local Rindler horizon.
\item  The reason that the undetermined constant $\Lambda$ appears in the field equation is now unambiguous: Since any scalar multiplied with the metric will give zero contribution on the boundary (recall that $\zeta^\nu n_\nu|_{\partial U}=0$) and a constant always has a vanishing covariant derivative, the method used here can only determine the field equations up to an arbitrary constant. In this sense considering the latter to be some short of cosmological constant does not seem to be  fully justified. 
\end{itemize}

\section{Field equations and diffeomorphism invariance: matter action}
\label{matter}

Let us now try to include in our framework the matter action  in order to check if the full field equations (\ref{f}) can be recovered from the variation of $S_{\rm tb}=S_{\rm B}+ S_{\rm M}$ as in~\cite{v1} or of $S_{\rm ts}=S_{\rm S}+ S_{\rm M}$ as in~\cite{paddy,paddy3,paddy2,v1}. We are going to argue here that the matter cannot be included in the problem. 

Recall that the matter action is built out of generally covariant scalar and therefore for any diffeomorphism that preserves the boundary ($\zeta^\mu n_\mu=0$) one has
\be
\label{smatt}
\delta_\zeta S_{\rm M}=-\frac{1}{2\kappa}\int_{\partial U} d^3 x\sqrt{|h|}\, L_{\rm M}\zeta^\mu n_\mu=0.
\ee
However, in \cite{paddy,paddy3,paddy2,v1} it was claimed that 
\bea
\label{varmatt1}
\delta_\zeta S_{\textrm{M}}&=&\int_U d^ 4 x \sqrt{-g}\,  T_{\mu\nu} \nabla^\mu \zeta^\nu,
\eea
which using the covariant conservation of the stress-energy tensor reduces to
\bea
\label{varmatt}
\delta_\zeta S_{\textrm{M}}&=&-\int_U d^4 x\sqrt{-g}\, \nabla^\mu(T_{\mu\nu} \zeta^\nu).
\eea
so it seems that one has a clear discrepancy in this case.

The way one one arrives to eq.~(\ref{varmatt1}) is by writing the variation of the matter action as
\bea
\label{varmatt2}
\delta_\zeta S_{\textrm{M}}&=&\int_U d^ 4 x \left[\sqrt{-g}\,  T_{\mu\nu} \delta_\zeta g^{\mu\nu}+\frac{\delta {\cal L}_{\rm M}}{\delta \psi}\delta_\zeta \psi\right],
\eea
where $\psi$ collectively denotes the matter fields, and using the fact the the matter fields satisfy their field equations to argue that the second term vanish. However, recall that the matter Lagrangian will generically contain terms of the sort $(\nabla \psi)^2$ and therefore will lead to surface terms. These normally would vanish under a standard variation with respect to $\psi$ due to the fact that $\psi$ is in this case kept fixed on the boundary, but would not vanish here since no such condition has to hold.
These terms are implicitly neglected when writing eq.~(\ref{varmatt1}) and this explains the discrepancy between eqs.~(\ref{smatt}) and (\ref{varmatt}). The straightforward validity of eq.~(\ref{smatt}) indicates clearly that eq.~(\ref{varmatt}) is incorrect. 

In order to illustrate this explicitly let us consider a simple scalar field $\phi$ as an example of a matter field: in this case the action of the diffeomorphism on the field will be $\delta_\zeta \phi=-\mathsterling_\zeta \phi=-\zeta^\mu \partial_\mu \phi$ which of course will not vanish on the boundary. 

Let us take the matter action of the generic form
\be
S^\phi_{\rm M}=\int_U d^4 x\sqrt{-g} L_{\rm M}(\phi,\partial_\mu \phi, g^{\mu\nu}),
\ee
under a diffeomorphism we shall get
\begin{widetext}
\bea
\label{varscfield}
\delta_\zeta S_{\textrm{M}}&=&
\int_U d^4 x \frac{\delta (\sqrt{-g} L_{\rm M})}{\delta g^{\mu\nu}} \delta_\zeta  g^{\mu\nu}+
\int_U d^4 x \sqrt{-g} \left[ \frac{\partial {L}_{\rm M}} {\partial \phi} \delta_\zeta \phi + \frac{\partial {L}_{\rm M}}{\partial \nabla^\mu \phi}  \delta_\zeta(\nabla^\mu \phi) \right]\\
&=& -\frac{1}{2} \int_U d^4 x \sqrt{-g} T_{\mu\nu} \delta_\zeta g^{\mu\nu}+
\int_U d^4 x \sqrt{-g} \left[ \frac{\partial {L}_{\rm M}}{\partial \phi} - \nabla^\mu\left(\frac{\partial {L}_{\rm M}}{\partial \nabla^\mu \phi}\right)\right] \delta_\zeta \phi+ \int_U d^ 4 x \sqrt{-g} \nabla^\mu \left(  \frac{\partial {L}_{\rm M}}{\partial \nabla^\mu \phi} \delta_\zeta \phi\right),\nn
\eea
\end{widetext}
where the stress-energy tensor of the scalar field is defined as
\be
\label{set}
T_{\mu\nu}^{\phi}=-\frac{2}{\sqrt{-g}}\frac{\delta (\sqrt{-g} L_{\rm M})}{\delta g^{\mu\nu}}.
\ee
Using the matter field equations  
\be\frac{\partial {L}_{\rm M}}{\partial \phi} - \nabla^\mu\left(\frac{\partial {L}_{\rm M}}{\partial \nabla^\mu \phi}\right)=0,\ee
and Stokes theorem one gets
\bea
\delta_\zeta S_{\textrm{M}}&=&  -\frac{1}{2} \int_U d^ 4 x \sqrt{-g} T^\phi_{\mu\nu} \delta_\zeta g^{\mu\nu}+\nn\\
&&+ \int_{\partial U} d^3 x \sqrt{|h|} \frac{\partial {L}_{\rm M}}{\partial \nabla^\mu \phi} n^\mu \delta_\zeta \phi
\eea
which for a diffeomorphism becomes
\bea
\delta_\zeta S_{\textrm{M}}&=&  - \int_U d^ 4 x \sqrt{-g} T^\phi_{\mu\nu} \nabla^\mu\zeta^\nu-\nn\\
&&- \int_{\partial U} d^3 x \sqrt{|h|} \frac{\partial {L}_{\rm M}}{\partial \nabla^\mu \phi} n^\mu \zeta^\nu \nabla_\nu \phi
\eea
so that in the end
\bea
\label{set2}
\delta_\zeta S_{\textrm{M}}&=& \int_U d^ 4 x \sqrt{-g} \nabla^\mu T^\phi_{\mu\nu} \zeta^\nu-\\
&&- \int_{\partial U} d^3 x \sqrt{|h|} \left( \frac{\partial {L}_{\rm M}}{\partial \nabla^\mu \phi} \nabla_\nu \phi+T^\phi_{\mu\nu} \right) n^\mu \zeta^\nu \nn
\eea
Let us focus on the surface integral in the above equation: introducing the so called canonical stress energy tensor (see \cite{wald}, p. 457) 
\be
T^c_{\mu\nu}=\frac{\partial {L}_{\rm M}}{\partial \partial^\mu \phi} \partial_\nu \phi - L_{\rm M} g_{\mu\nu}
\ee
one can rewrite the argument of the surface integral as
\be
\left (T^c_{\mu\nu}- L_{\rm M} g_{\mu\nu}+T^\phi_{\mu\nu} \right) n^\mu\zeta^\nu=\left (T^c_{\mu\nu}+T^\phi_{\mu\nu}\right) n^\mu\zeta^\nu
\ee
where we have used that $\zeta^\mu n_\mu=0$.  However for any scalar field the canonical SET and the standard one coincide but have the opposite sign and hence the integrand simply cancels. We remain then with the last integral over the bulk in (\ref{set2}) which is obviously zero if the stress energy tensor is conserved.
Hence in the end we recover, as expected, that the matter action is invariant under diffemorphisms that preserve the boundary whenever the matter field equations and the conservation of the SET hold. Alternatively one can use the invariance of the action and the matter field equations to recover the conservation of the SET even without presuming that the generator of the diffeomorphism $\zeta$ identically vanishes on the boundary.

As we said the validity of eq.~(\ref{smatt}) is anyway unquestionable and we performed the last calculation just to clarify what is the inconsistent assumption that leads to eq.~(\ref{varmatt}). This turns out to be the fact that part of the surface term where not taken properly into account in \cite{paddy,paddy3,paddy2,v1} and are of course crucial since the whole discussion is based on integrals over the boundary. We can now easily realize that one cannot derive field equations including matter by requiring the actions $S_{\rm tb}$ or $S_{\rm ts}$ to be invariant under diffeomorphisms that are normal to a null boundary  since the matter action is itself an invariant.

 Another point that should be stressed is the following. If matter is indeed added to the problem one cannot even require $S_{\rm tb}$ or $S_{\rm ts}$ to be an invariant under the special class of diffeomorphisms discussed in the previous sections. The action that leads to the field equations should be invariant under diffeomorphisms that leave the boundary untouched. This does not only regard the topological characteristics of the boundary --- whether it is mapped to itself --- but is also relevant to the values of the fields there. When the metric is not the only field present, even if one assumes that $\zeta^\mu$ is tangential to the boundary, condition (\ref{cond1}) is not enough to guarantee that $U$ is mapped on itself and at the same time that all the field remain unaffected on the boundary. The matter fields will be affected even if condition (\ref{cond1}) holds. In the presence of matter fields only the standard diffeomorphisms that vanish on the boundary really map $U$ on $U$ and at the same time leave all fields unaffected on the boundary, and it is obvious that $S_{\rm tb}$ is invariant under them, as it should be. Of course, as it was discussed previously, considering diffeomorphisms that vanish on the boundary cannot lead to field equations.

One can of course avoid the problem with the matter action discussed above by demoting the diffeomorphism to a restricted metric variation. This was recently also proposed in a revised version \cite{PadRev} of \cite{paddy}. In this sense $\zeta^\mu$ will no longer be a vector generating a diffeomorphism as it was claimed in \cite{paddy,paddy3,paddy2} but one would just assume that the metric is varied in such a way that $\delta g^{\mu\nu}=\nabla^\mu \zeta^\nu+\nabla^\nu \zeta^\mu$  always holds and $\zeta^\mu$ vanishes on the bulk and is a null vector on the horizon. In this case the variation would not affect the matter field and eq.~(\ref{varmatt}) would indeed hold. Thus if one demands the action
$ S_{\rm tot}=S_{\rm S}+S_{\rm M}$
to be stationary under such variations then field equation can be derived. 
However, it is a fact that in this case the field equations do no longer come as a consequence of a basic principle like  
diffeomorphism invariance, but through a very restricted form of a  
metric variation (which is the  standard way to derive them anyway).  
This diminishes, in our opinion, the appeal of this approach although  
there are proposals which are trying to explain why such specific  
metric variations should play a  special role (see {\it e.g.~}discussion in ref. \cite{PadRev}). 

\section{Conclusions}
\label{concl}

We have discussed a recent claim that field equations can be derived by requiring the invariance of an action built out of the sum of the surface part of the gravitational action and the matter action under a specific class of diffeomorphisms. After examining thoroughly the procedure that one can follow to justify such an invariance we showed that this claim is incorrect, due to the fact that the matter action is built out of a generally covariant scalar and, therefore, remains unchanged under such diffeomorphisms. 

 However, in the case of pure gravity the claim still holds and reveals an important aspect of gravitational actions that lead to second order equations. One can use diffeomorphism invariance to derive field equations from both the bulk part and the surface part of the gravitational action. This is due to the fact that when the metric is the only field present, there is a special class of diffeomorphisms, different than those that vanish on the boundary and are usually mentioned in textbooks,  that map a compact region of the manifold to itself and at the same time leave the metric unaffected there. This approach demonstrates the importance of surface terms, since these are the terms that led to the field equations. In this setting the surface term of the gravitational action seems to contain all the information about spacetime that are contained in the bulk action. The fact that the approach does not hold in the presence of matter does not invalidate the specific nature of the surface term in the gravitational action. It merely implies that matter does not have a similar behaviour and cannot be treated on the same grounds.

One final comment we would like to make is about the use of the term holography in \cite{paddy,paddy3,paddy2}. Even though, we argued for a special behaviour of gravity and it is a fact that surface terms turn out to be crucial and sufficient to derive the field equations in the absence of matter, this does not imply a holographic behaviour in the literal meaning of the word. To argue for the validity of the derived field equations throughout spacetime, one is always obliged to use the fact that each point of spacetime can be assumed to be a point of a local Rindler horizon. Therefore, it is not exact to say that the gravitational degrees of freedom of a single region reside in its boundary surface. 

%%%%%%%%%%%%%%%%%%%%%%%%%%%%%
\section*{Acknowledgements}
The authors are indebted to David Mattingly, Martin Charest, Sijie Gao, Hongbao Zhang and Sebastiano Sonego for their contribution in pointing out inconsistencies in a previous version of this paper. They also wish to thank Thanu Padmanabhan, John Miller and Enrico Barausse for useful discussions.

\end{document}